\documentclass{article}
\usepackage{url}
\usepackage{amsmath}
\usepackage{amsthm}
\usepackage{amssymb}
\usepackage{amstext}

\newtheorem{lemma}{Lemma}
\newtheorem{definition}{Definition}

\newtheorem{corollary}{Corollary}
\newtheorem{theorem}{Theorem}
\newtheorem{proposition}{Proposition}

\newcommand*{\cB}{\mathcal{B}}
\newcommand*{\cS}{\mathcal{S}}
\newcommand*{\cT}{\mathcal{T}}
\newcommand*{\cX}{\mathcal{X}}
\newcommand*{\cY}{\mathcal{Y}}
\newcommand*{\bbR}{\mathbb{R}}
\newcommand*{\bbN}{\mathbb{N}}
\newcommand*{\bx}{\mathbf{x}}
\newcommand*{\by}{\mathbf{y}}
\newcommand*{\Xb}{\bar{X}}
\newcommand*{\cXb}{\bar{\cX}}
\newcommand*{\Yb}{\bar{Y}}

\DeclareMathOperator{\ExpE}{E}
\DeclareMathOperator{\supp}{supp}

\newcommand*{\Hmax}{H_{\mathrm{max}}}
\newcommand*{\Hmin}{H_{\mathrm{min}}}
\newcommand*{\eps}{\varepsilon}
\newcommand*{\enc}{\mathrm{enc}}

\newcommand*{\ext}{\mathrm{ext}}

\title{On the Randomness of Independent Experiments}
\author{Thomas Holenstein\thanks{Department of Computer
    Science; ETH Z\"urich; Switzerland; {\tt
      thomahol@inf.ethz.ch}} \addtocounter{footnote}{2} and Renato Renner\thanks{Department of
    Applied Mathematics and Theoretical Physics; University of
    Cambridge; United Kingdom; {\tt r.renner@damtp.cam.ac.uk}}}

\begin{document}

\maketitle

\begin{abstract}  
  Given a probability distribution $P_X$, what is the minimum amount
  of bits needed to store a value $x$ sampled according to $P_X$, such
  that $x$ can later be recovered (except with some small probability
  $\eps$)? Or, what is the maximum amount of uniform randomness that
  can be extracted from~$x$? Answering these and similar
  information-theoretic questions typically boils down to computing
  so-called \emph{smooth entropies}.  In this paper, we derive
  explicit and almost tight bounds on the smooth entropies of $n$-fold
  product distributions $P_X^n$.
\end{abstract}

\section{Introduction}

\subsection{Smooth min- and max-entropy}
Smooth min- and max-entropy has been introduced
in~\cite{RenWol04a,RenWol05b} as a generalization of Shannon entropy.
Similarly to Shannon entropy, smooth min- and max-entropy can be used
to analyze information-processing tasks such as data compression.
However, in contrast to Shannon entropy, which usually only makes
sense in an asymptotic setting (where an underlying random experiment
is repeated many times), smooth entropies can also be used in the 
non-asymptotic case.

We start by quickly reviewing the relevant definitions. For the
following, let $X$ and $Y$ be random variables with range $\cX$ and
$\cY$, respectively, and joint distribution $P_{XY}$.  Moreover, for
$\eps \geq 0$, let $\cB^\eps(P_{XY})$ be the $\eps$-ball of
nonnegative functions around $P_{X Y}$, i.e., the set of functions
$Q_{X Y}: \, \cX \times \cY \to \bbR^+$ such that $\|P_{X Y} - Q_{X
  Y}\|_1 \leq \eps$, where $\| \cdot \|_1$ denotes the $L_1$-norm.

\begin{definition}
  The \emph{$\eps$-smooth max-entropy} of $X$ given $Y$
  is\footnote{We use~$\log(\cdot)$ to denote the binary logarithm.}
  \[
    \Hmax^\eps(X|Y) 
  := 
    \min_{Q_{X Y} \in \cB^\eps(P_{XY})} \max_{y \in \cY} \log \bigl|\supp Q_{X Y}(\cdot, y) \bigr| 
  \]
  where $\supp Q_{X Y}(\cdot, y)$ denotes the support of the function
  $Q_{X Y}(\cdot, y): \, x \mapsto Q_{X Y}(x,y)$.  
  
  The \emph{$\eps$-smooth min-entropy} of $X$ given $Y$ is
  \[
    \Hmin^\eps(X|Y)
  :=
    \max_{Q_{X Y} \in \cB^\eps(P_{XY})} \min_{y \in \supp P_Y} \min_{x \in \cX} \log \frac{P_Y(y)}{Q_{X Y}(x,y)} 
  \]
  where $P_Y$ denotes the marginal distribution of $P_{XY}$.
\end{definition}

The following two statements proven in~\cite{RenWol05b} imply that the
smooth min- and max-entropies have a (non-asymptotic) operational
interpretation. For example, the smooth max-entropy characterizes
\emph{data compression}. More precisely, for $(x,y)$ chosen according
to $P_{X Y}$, it quantifies the minimum space needed to store $x$ such
that, with the help of $y$, the value $x$ can later be retrieved
(except with probability at most $\eps$).

\begin{proposition}\label{prop:hmax}
  Let $\ell_{\enc}^\eps(X|Y)$ be the minimum number $\ell$ such that
  \[
    \Pr_{(x,y) \leftarrow P_{X Y}}\bigl[d(e(x),y) \neq x\bigr] \leq \eps 
  \]
  for some encoding function $e : \, \cX \to \{0,1\}^{\ell}$ and some
  decoding function $d: \, \{0,1\}^\ell \times \cY \to \cX$.  Then,
  for any $0 \leq \eps' < \eps$,
  \[
    \Hmax^\eps(X|Y) \leq \ell_{\enc}^\eps(X|Y) \leq \Hmax^{\eps'}(X|Y) + \log(1/(\eps-\eps'))+1 \ .
  \]
\end{proposition}

Similarly, the smooth min-entropy characterizes \emph{randomness
  extraction}. That is, for $(x,y)$ chosen according to $P_{X Y}$, it
corresponds to the maximum number of bits that can be computed from
$x$ such that these bits are uniformly distributed and independent of
$y$ (except with probability $\eps$).

\begin{proposition}\label{prop:hmin}
  Let $\ell_{\ext}^\eps(X|Y)$ be the maximum number $\ell$ such that
  \[
     \frac{1}{2}\bigl\| P_{h(X) Y} - P_U \times P_Y \bigr\|_1 \leq \eps
  \]
  for some extraction function $h: \, \cX \to \{0,1\}^\ell$, where $P_U$
  is the uniform distribution on $\{0,1\}^\ell$.  Then, for any $0
  \leq \eps' < \eps$,
  \[
    \Hmin^{\eps'}(X|Y) - 2\log(1/(\eps-\eps')) \leq \ell_{\ext}^\eps(X|Y) 
    \leq \Hmin^{\eps}(X|Y)  \ .
  \]
\end{proposition}

While, by the above propositions, smooth entropies are directly
related to data compression and randomness extraction, they are also
useful for the characterization of a variety of other tasks in
communication theory and cryptography~\cite{RenWol05b,ReWoWu06}.
Moreover, they can be generalized to quantum states~\cite{Renner05}.

\subsection{Contributions of this paper}
In this paper, we are concerned with the explicit computation of
smooth entropies for the case of a finite number of independently
repeated experiments. More precisely, we derive the following bounds
on the smooth min- and max-entropies $\Hmin^\eps(X^n|Y^n)$ and
$\Hmax^\eps(X^n|Y^n)$ of an $n$-fold product distribution $P_{X^n Y^n}
:= P_{X_1 Y_1}\times\dots\times P_{X_nY_n}$.
\begin{theorem} \label{thm:classprodentr} 
  Let $P_{X^n Y^n} := P_{X_1Y_1}\dots P_{X_nY_n}$ be
  a probability distribution over~$\cX^n\times\cY^n$.  For any $\delta
  \geq 0$
  \begin{align*}
    \Hmax^\eps(X^n|Y^n)  
  & \leq
    H(X^n|Y^n) + n\delta \ ,
  \\
    \Hmin^\eps(X^n|Y^n)  
  & \geq
    H(X^n|Y^n) - n\delta \ ,
  \end{align*}
  where $\eps = 2^{-\frac{n \delta^2}{2 \log^2(|\cX| + 3)}}$. 
% $\delta :=\sqrt{\frac{2 \log(1/\eps)}{n}} \log(|\cX|+3) $.
\end{theorem}

Let~$(\bx,\by)$ be chosen according to the $n$-fold product
distribution $P_{X^nY^n} = P_{X_1Y_1}\dots P_{X_nY_n}$.  The well
known asymptotic equipartition property states that
\begin{align*}
  \lim_{n\to\infty}\bigl(\Pr_{\bx,\by}[P_{X^n|Y^n}(\bx,\by) \in
  2^{-H(X^n|Y^n)\pm n\delta}]\bigr) = 1
\end{align*}
for every~$\delta>0$.  The main step
in order to prove Theorem~\ref{thm:classprodentr} is to give the
following quantitative bound on this convergence.
\begin{theorem} \label{thm:probchernoff}
  Let $P_{X^n Y^n} := P_{X_1 Y_1}\dots P_{X_n Y_n}$ be a probability
  distribution over~$\cX^n\times\cY^n$.  Then, for any $\delta \in [0,
  \log(|\cX|)]$ and $(\bx, \by)$ chosen according to $P_{X^n Y^n}$,
  \[
    \Pr_{\bx,\by} \bigl[
        -\log(P_{X^n|Y^n}(\bx,\by)) 
        \geq  H(X^n|Y^n) + n\delta \bigr] 
  \leq 
    \eps \ ,
  \] 
  and, similarly,
  \[
    \Pr_{\bx,\by} \bigl[
        - \log(P_{X^n|Y^n}(\bx, \by)) 
        \leq  H(X^n|Y^n) - n\delta \bigr] 
  \leq 
    \eps  \ ,
  \]   
  where $\eps = 2^{- \frac{n \delta^2}{2 \log^2(|\cX|+3)} }$.
\end{theorem}

We prove the theorems in Section~\ref{sec:smoothprod}.  In
Section~\ref{sec:tightness} we show that both these theorems are
almost tight (cf.~Theorems~\ref{thm:optimalityOne}
and~\ref{thm:optimalityTwo}).

\subsection{Related work and proof technique}

A bound as in Theorem~\ref{thm:probchernoff} can be obtained in
simpler ways than the one we use.  However, we only know of simpler
arguments which yield quantitatively weaker bounds.

We sketch two such arguments.  The first argument (which only seems to
work in case the distributions~$P_{X_iY_i}$ are identical) goes as
follows: let~$Q_{\bx|\by}$ be the frequency distribution (i.e., the
type) of the pair~$(\bx,\by) = ((x_1,y_1),\ldots,(x_n,y_n))$.  It is
well known that $D(Q_{\bx|\by}\|P_{X|Y})$ is small with high
probability, and an explicit bound can be given \cite[Theorem
12.2.1]{CovTho91}.  The Csisz\'ar-Kullback-Pinsker inequality
\cite[Lemma 12.6.2]{CovTho91} then shows that $\|Q_{\bx|\by} -
P_{X|Y}\|_1$ is small with high probability, which in turn bounds
$|H(Q_{\bx|\by})-H(P_{X|Y})|$ by the Fano inequality (more concretely
\cite[Theorem 16.3.2]{CovTho91}), and a bound on
$|H(Q_{\bx|\by})-H(P_{X|Y})|$ is exactly what we want.  This argument
shows that the probabilities in Theorem \ref{thm:probchernoff} are at
most
\begin{align*}
  2^{-n\cdot \Theta\left(\frac{\delta^2}{\log^2(|\cX|/\delta)} - 
    |\cX|\frac{\log(n+1)}{n}\right)}\,.
\end{align*}
The major problem with this bound is that it is only useful if~$n\gg
|\cX|$, but additionally the term~$\log^2(|\cX|/\delta)$ is not tight:
it can be strengthened to~$\log^2(|\cX|)$ as our proof shows (which is
interesting if~$\delta \ll \frac{1}{|\cX|}$).

A different argument is used in \cite{ImLeLu89}.  There, the Hoeffding
bound is applied on the sum of the independent random
variables~$\log(1/P_{X|Y}(X_i,Y_i))$.  Unfortunately, the Hoeffding
bound can only be applied if the random variables have a bounded
range, and thus one ignores occurrences where this random variable is
very large (which happens with some small probability).  Hence, this
technique only gives a bound on the second probability in
Theorem~\ref{thm:probchernoff}, and this bound is
\begin{align*}
  2^{-n\cdot \Theta\left(\frac{\delta^2}{\log^2(|\cX|/\delta)}\right)}.
\end{align*}
As above, this is not tight in case~$\delta \ll \frac{1}{|\cX|}$.

Instead of using Hoeffding's bound we directly use Chernoff's argument
\cite{Cherno53} which states that an upper bound on~$\inf_t
e^{-at}M(t)$ for every~$a$ suffices for our purpose, where~$M(t)$ is
the moment generating function of the random variable
$\log(1/P_{X|Y}(X,Y))$.  In order to make the presentation simpler we
do not use Chernoff's theorems explicitly, but instead give the
complete proof.

\section{Smooth min- and max-entropy of products} \label{sec:smoothprod}

\subsection{Typical sequences and their probabilities}
\begin{lemma} \label{lem:Etbound}
  Let $P_{X Y}$ be a probability distribution on $\cX \times \cY$.
  Then, for any $t \in \bbR$ with $|t| \leq \frac{1}{\log(|\cX|+3)}$,
  \[
    \log\Bigl(\ExpE_{x,y}\bigl[P_{X|Y}(x,y)^{-t}\bigr]  \Bigr)
  \leq 
    t H(X|Y) + \tfrac{1}{2} t^2 \log^2(|\cX|+3) \ ,
  \]
  where the expectation is taken over pairs $(x,y)$ chosen according
  to $P_{X Y}$.
\end{lemma}

\begin{proof}
  For any $t \in \bbR$, let $r_t$ be the function on the open interval
  $(0, \infty)$ defined by
  \begin{equation} \label{eq:rtdef}
    r_t(z) := z^t - t \ln(z) -1 \ .
  \end{equation}
  We will use several properties of this function proven in
  Appendix~\ref{sec:rt}.

  For any $x \in \cX$ and $y \in \cY$, let $p_{x,y}:=P_{X|Y}(x,y)$.
  If $p_{x,y} > 0$ then
  \[
    p_{x,y}^{-t}
  =
    r_t\bigl(\tfrac{1}{p_{x,y}}\bigr) + t \ln\bigl(\tfrac{1}{p_{x,y}}\bigr) + 1
  \leq
    r_t\bigl(\tfrac{1}{p_{x,y}}+3\bigr) + t \ln\bigl(\tfrac{1}{p_{x,y}}\bigr) 
    + 1 \ ,
  \]
  where the inequality holds because $r_t$ is monotonically increasing
  on the interval $[1, \infty)$ (Lemma~\ref{lem:rtincr}) and
  $\frac{1}{p_{x,y}} = \frac{P_Y(y)}{P_{X Y}(x,y)} \geq 1$.  Because
  $\frac{1}{p_{x,y}} + 3 \in [4, \infty)$ and because $r_t$ is concave
  on this interval (Lemma~\ref{lem:rtconc} which can be applied
  because $t \in [-\frac{1}{2}, \frac{1}{2}]$), Jensen's inequality
  leads to
  \[  
  \begin{split}
    \ExpE_{x,y}\bigl[p_{x,y}^{-t}\bigr]
  & \leq
    \ExpE_{x,y}\Bigl[r_t\bigl(\tfrac{1}{p_{x,y}} + 3\bigr) \Bigr] 
    + t \ExpE_{x,y} \bigl[\ln\bigl(\tfrac{1}{p_{x,y}}\bigr)\Bigr] + 1 \\
  & \leq
    r_t\Bigl(\ExpE_{x,y}\bigl[\tfrac{1}{p_{x,y}}+3\bigr]\Bigr) 
    + t \ln(2) \ExpE_{x,y} \Bigl[\log\bigl(\tfrac{1}{p_{x,y}}\bigr)\Bigr] + 1 
    \ .
  \end{split}
  \]
  Because $\ExpE_{x,y}[\frac{1}{p_{x,y}}] = \sum_{x,y} P_{X Y}(x,y)
  \frac{P_{Y}(y)}{P_{X Y}(x,y)} = |\cX|$ and
  $\ExpE_{x,y}[\log(\frac{1}{p_{x,y}})] = H(X|Y)$, we obtain
  \[
    \ExpE_{x,y}\bigl[p_{x,y}^{-t}\bigr]
  \leq
    r_t(|\cX|+3) + t \ln(2) H(X|Y) + 1 \ .
  \]
  Furthermore, because $\log(a) \leq \frac{1}{\ln(2)}(a-1)$,
  \[
    \log\Bigl(\ExpE_{x,y}\bigl[p_{x,y}^{-t}\bigr]\Bigr)
  \leq
    \tfrac{1}{\ln(2)} r_t(|\cX|+3) + t H(X|Y) \ .
  \]
  Finally, together with Lemma~\ref{lem:rtbound}, since $|t| \leq
  \frac{1}{\log(|\cX|+3)}$, we conclude
  \[
    \log\Bigl(\ExpE_{x,y}\bigl[p_{x,y}^{-t}\bigr]\Bigr)
  \leq
    \bigl(\tfrac{1}{\ln(2)} - 1 \bigr) t^2 \log^2(|\cX|+3) + t H(X|Y) \ .
  \]
  The assertion follows because $\frac{1}{\ln(2)} - 1 \leq
  \frac{1}{2}$.
\end{proof}

\newcommand*{\entrm}{\gamma}

\begin{lemma} \label{lem:exptZbound}
  Let $P_{X Y}$ be a probability distribution and let $\entrm$ be the
  function on $\cX \times \cY$ defined by
  \[
    \entrm(x,y) := - \log(P_{X|Y}(x,y)) - H(X|Y) \ .
  \]
  Then, for any $t \in \bbR$ with $|t| \leq \frac{1}{\log(|\cX|+3)}$,
  \[
    \ExpE_{x,y}\bigl[2^{t \entrm(x,y)}\bigr] 
  \leq 
    2^{\frac{1}{2} t^2 \log^2(|\cX|+3)} \ .
  \]
\end{lemma}

\begin{proof}
  The assertion follows directly from Lemma~\ref{lem:Etbound}, that is,
  \[
  \begin{split}
    \ExpE_{x,y}\bigl[2^{t \entrm(x,y)}\bigr] 
  & =
    2^{-t H(X|Y)} \ExpE_{x,y}\bigl[P_{X|Y}(x,y)^{-t}\bigr] \\
  & \leq
    2^{-t H(X|Y)} 
    \cdot  2^{t H(X|Y) + \frac{1}{2} t^2 \log^2(|\cX|+3)} \ .  \qedhere
  \end{split}
  \]
\end{proof}

\begin{proof}[Proof (of Theorem~\ref{thm:probchernoff})]
  Let $\bx=(x_1, \ldots, x_n)$, $\by=(y_1, \ldots, y_n)$, and let
  $\entrm_i$ be the function defined in Lemma~\ref{lem:exptZbound} for
  the probability distribution $P_{X_i Y_i}$. Then
  \begin{equation} \label{eq:zxysum}
    \sum_{i=1}^n \entrm_i(x_i, y_i)
  =
    - \log(P_{X^n|Y^n}(\bx, \by)) 
    - H(X^n|Y^n) \ .
  \end{equation}
  Using Markov's inequality, for any $t > 0$,
  \begin{equation} \label{eq:HsMarkov}
    \Pr_{\bx,\by}\Bigl[\sum_{i=1}^n \entrm_i(x_i, y_i) \geq n \delta\Bigr]
  =
    \Pr_{\bx,\by}
      \bigl[2^{t \sum_{i=1}^n \entrm_i(x_i, y_i)} \geq 2^{t n \delta}\bigr] \\
  \leq
    \frac{\ExpE_{\bx,\by}
      \bigl[2^{t \sum_{i=1}^n \entrm_i(x_i, y_i)}\bigr]}{2^{t n \delta}} \ .
  \end{equation}
  Moreover, because the pairs $(x_i, y_i)$ are chosen independently,
  \begin{align*}
    \ExpE_{\bx,\by}\Bigl[ 2^{t \sum_{i=1}^n \entrm_i(x_i, y_i)} \Bigr]
  &=
    \ExpE_{\bx,\by}\Bigr[ \prod_{i=1}^n 2^{t \entrm_i(x_i, y_i)} \Bigr]\\
  &=
    \prod_{i=1}^n  \ExpE_{x_i, y_i}\bigr[2^{t \entrm_i(x_i, y_i)} \bigr] 
  \leq  
    \bigl( 2^{\frac{1}{2} t^2 \log^2(|\cX|+3)} \bigr)^n
  \end{align*}
  where the inequality follows from Lemma~\ref{lem:exptZbound}, for
  any $|t| \leq \frac{1}{\log(|\cX|+3)}$.  Combining this
  with~\eqref{eq:HsMarkov} gives
  \[
    \Pr_{\bx,\by}\Bigl[\sum_{i=1}^n \entrm_i(x_i, y_i) \geq n \delta\Bigr]
  \leq
    2^{\frac{1}{2} n t^2 \log^2(|\cX|+3) - t n \delta} \ .
  \]
  With $t := \frac{\delta}{\log^2(|\cX|+3)}$ (note that $t \leq
  \frac{1}{\log(|\cX| +3)}$ because $\delta \leq \log(|\cX|)$), we
  conclude
  \[
    \Pr_{\bx,\by}\Bigl[\sum_{i=1}^n \entrm_i(x_i, y_i) \geq n \delta\Bigr]
  \leq
    2^{- \frac{n \delta^2}{2 \log^2(|\cX|+3)} } \ .
  \]
  The first inequality of the theorem then follows
  from~\eqref{eq:zxysum}.
 
  Similarly, if $t < 0$,
  \[
    \Pr_{\bx,\by}\Bigl[\sum_{i=1}^n \entrm_i(x_i, y_i) \leq -n \delta\Bigr]
  =
    \Pr_{\bx,\by}
      \Bigl[2^{t \sum_{i=1}^n \entrm_i(x_i, y_i)} \geq 2^{- t n \delta}\Bigr] \\
  \leq
    \frac{\ExpE_{\bx,\by}
      \bigl[2^{t \sum_{i=1}^n \entrm_i(x_i, y_i)}\bigr]}{2^{- t n \delta}}
   \ ,
  \]
  and thus
  \[
    \Pr_{\bx,\by}\Bigl[\sum_{i=1}^n \entrm_i(x_i, y_i) \leq -n \delta\Bigr]
  \leq
    2^{\frac{1}{2} n t^2 \log^2(|\cX|+3) + t n \delta} \ .
  \]
  The second inequality follows with $t :=
  -\frac{\delta}{\log^2(|\cX|+3)}$.
\end{proof}

\subsection{Asymptotic equality of smooth entropy and Shannon entropy}

\begin{proof}[Proof (of Theorem~\ref{thm:classprodentr})]
  We first prove the bound on the max-entropy $\Hmax^\eps(X^n|Y^n)$.
  For any $\by \in \cY^n$ with $P_{Y^n}(\by) > 0$, let $\cXb_{\by}$ be
  the set of all $n$-tuples $\bx \in \cX^n$ such that
  \[
    -\log(P_{X^n|Y^n}(\bx,\by))
%   =
%     \log(\frac{1}{P_{X|Y}(x_1, y_1) \cdots P_{X|Y}(x_n, y_n)}) 
  \leq 
     H(X^n|Y^n) + n\delta \ .
  \]
  Furthermore, let $P_{\Xb^n \Yb^n}$ be the nonnegative function on
  $\cX^n \times \cY^n$ defined by
  \begin{equation} \label{eq:PXbYbdef}
    P_{\Xb^n \Yb^n}(\bx, \by)
  =
    \begin{cases}
      P_{X^n Y^n}(\bx, \by) & \text{if $\bx \in \cXb_{\by}$} \\
      0 & \text{otherwise.}
    \end{cases}
  \end{equation}
  We can assume without loss of generality that $\delta \leq
  \log(|\cX|)$ (otherwise, the statement is trivial). Hence, by the
  first inequality of Theorem~\ref{thm:probchernoff},
  $\Pr_{\bx,\by}[\bx \notin \cXb_{\by}] \leq \eps$, which implies
  \begin{align}\label{eq:4}
    \bigl\|P_{X^n Y^n} - P_{\Xb^n \Yb^n} \bigr\|_1 \leq \eps\, .
  \end{align}
  For any fixed $\by:=(y_1, \ldots, y_n) \in \cY^n$ with $P_{Y^n}(\by)
  > 0$,
  \[
    1 
  \geq 
    \sum_{\bx \in \cXb_{\by}} 
      P_{X_1|Y_1}(x_1,y_1) \cdots  P_{X_n|Y_n}(x_n,y_n)
  \geq 
    |\cXb_{\by}| 2^{- H(X^n|Y^n) - n\delta} \ ,
  \]
  where the second inequality follows from the definition of the set
  $\cXb_{\by}$. Consequently, we have $|\cXb_{\by}| \leq 2^{H(X^n|Y^n)
    + n\delta}$.  Moreover, by the definition of $P_{\Xb^n \Yb^n}$,
  the support of the function $\bx \mapsto P_{\Xb^n \Yb^n}(\bx,\by)$
  is contained in $\cXb_{\by}$.  By the definition of
  max-entropy and (\ref{eq:4})
  \begin{align*}
    \Hmax^\eps(X^n|Y^n)
    \leq
    \max_{\by\in\cY^n} \log \bigl|\supp P_{\Xb^n \Yb^n}(\cdot, \by)\bigr|
    \leq
    H(X^n|Y^n) + n\delta\,.
  \end{align*}
  
  To prove the bound on the min-entropy $\Hmin^\eps(P_{X^n
    Y^n}|P_{Y^n})$, let $\cXb_{\by}$, for any $\by \in \cY^n$ with
  $P_{Y^n}(\by) > 0$, be the set of $n$-tuples $\bx \in \cX^n$ such
  that
  \[
    - \log(P_{X^n|Y^n}(\bx,\by)) 
  \geq 
     H(X^n|Y^n) - n\delta \ ,
  \]
  and let again $P_{\Xb^n \Yb^n}$ be defined by~\eqref{eq:PXbYbdef}.
  By the second inequality of Theorem~\ref{thm:probchernoff},
  $\Pr_{\bx,\by}[\bx \notin \cXb_{\by}] \leq \eps$, which gives
  \begin{equation*} 
    \bigl\| P_{X^nY^n}- P_{\Xb^n\Yb^n} \bigr\|
  \leq
    \eps \ .
  \end{equation*}
  Thus, by the definition of min-entropy,
  \begin{align*}
    \Hmin^\eps(X^n|Y^n)
    & \geq
    \min_{\by \in \supp P_{Y^n}}\min_{\bx \in \cX^n}
       \log\frac{P_{Y^n}(\by)}{P_{\Xb^n\Yb^n}(\bx,\by)} \\
    & \geq H(X^n|Y^n)-n\delta.
  \end{align*}
  where the second inequality follows from the definition of the set
  $\cXb_{\by}$. 
\end{proof}

%Because the min-entropy $\Hmin(P_{X^n Y^n}|P_{Y^n})$ cannot be larger
%than the max-entropy $\Hmax(P_{X^n Y^n}|P_{Y^n})$,
%Theorem~\ref{thm:classprodentr} implies that
%\begin{equation} \label{eq:productapprox}
%  \tfrac{1}{n} \Hmin^\eps(P_{X^n Y^n}|P_{Y^n}) 
%\approx 
%  \tfrac{1}{n} \Hmax^\eps(P_{X^n Y^n}|P_{Y^n}) 
%\approx 
%\tfrac{1}{n} H(X^n|Y^n) \ ,
%\end{equation}
%where, asymptotically, for increasing $n$, the approximation becomes
%an equality.

%That is, for any distribution of the form $P_{X^n Y^n} =
%\prod_{i=1}^n P_{X_i Y_i}$, the
%approximation~\eqref{eq:productapprox} still holds.

%Note also that $\Hmin(X|Y) \leq H_{\max}(X|Y)$.
%Theorem~\ref{thm:classprodentr} does gives upper and lower bounds for
%both min- and max-entropies. [ ....]

\section{On the Tightness of the Bounds}\label{sec:tightness}
In this section, we show that Theorems~\ref{thm:classprodentr}
and~\ref{thm:probchernoff} are almost tight.  For this, we construct
one particular family of distributions for which these theorems cannot
be strengthened much.

Consider the set $\cX = \{0,\ldots,|\cX|-1\}$, where~$|\cX| \geq 3$.
We set
\begin{align}\label{eq:3}
  P_X(x) := \begin{cases} 
    \frac{1}{2} & x = 0\\
    \frac{1}{2(|\cX|-1)} & \text{otherwise.}
  \end{cases}
\end{align}
An explicit calculation gives 
\begin{align}\label{eq:6}
  H(X) = 1 + \frac{1}{2}\log(|\cX|-1).
\end{align}
For a tuple $\bx := (x_1, \ldots, x_n)$ let $z(\bx) := \bigl| \{i \mid
x_i = 0\}\bigr|$ be the number of zeros in~$\bx$.  Then,
\begin{align}
  (- \log(P_{X^n}(\bx))) - nH(X) &= 
  z(\bx) + (n-z(\bx)) (1+\log(|\cX|-1)) - n H(X)\nonumber\\
  &= \Bigl(\frac{n}{2}-z(\bx)\Bigr)\log(|\cX|-1).\label{eq:5}
\end{align}
Further, we note that~$z(\bx)$ is binomially distributed
with~$p=\frac12$.

We first show that Theorem~\ref{thm:probchernoff} is almost tight.
For this we use a lower bound on a partial sum over binomial
coefficients which is given in Appendix~\ref{sec:binom}
(Lemma~\ref{lem:bernoullisumestimate}).

\begin{theorem} \label{thm:optimalityOne}
  For any~$\cX$ with~$|\cX|\geq3$ there exists a distribution~$P_X$
  over~$\cX$ such that for $n \geq 12$, $0 \leq \delta \leq
  \frac{\log(|\cX|-1)}{12}$, and for $\bx$~chosen according to
  $P_{X^n} := (P_{X})^n$,
  \[
  \Pr_{\bx} 
  \bigl[-\log(P_{X^n}(\bx)) \geq  n \bigl(H(X) + \delta\bigr) \bigr] 
  >
  \frac{1}{110}2^{-\frac{12n\delta^2}{\log^2(|\cX|-1)}}
  \] 
  and, similarly,
  \[
    \Pr_{\bx} \bigl[
        - \log(P_{X^n}(\bx))
        \leq n \bigl(H(X) - \delta\bigr) \bigr] 
  >
  \frac{1}{110}2^{-\frac{12n\delta^2}{\log^2(|\cX|-1)}}\,.
  \]   
\end{theorem}
\begin{proof}
%  For an arbitrary element $x^* \in \cX$ consider the distribution 
%  \begin{align*}
%    P_X(x) := \begin{cases} 
%      \frac{1}{2} & x = x^*\\
%      \frac{1}{2(|\cX|-1)} & \text{otherwise,}
%    \end{cases}
%  \end{align*}
%  for which $H(X) = 1 + \frac{1}{2}\log(|\cX|-1)$.  Let now $\bx :=
%  (x_1, \ldots, x_n)$ be chosen according to $P_{X^n} := (P_{X})^n$,
%  and let $t := \bigl| \{i \mid x_i\neq x^*\}\bigr|$ be the number of
%  occurrences of elements different from~$x^*$ in $\bx$.  Clearly, $t$
%  is binomially distributed with $p=\frac{1}{2}$, and also
%  \begin{align*}
%    P_{X^n}(\bx) &= 
%    \frac{1}{2^{n-t}} \frac{1}{\bigl(2(|\cX|-1)\bigr)^{t}} \\
%    & = \frac{1}{2^n} \frac{1}{(|\cX| - 1)^{t}}.
%  \end{align*}
%  In order to obtain the first bound we manipulate the difference
%  \begin{align*}
%    (- \log(P_{X^n}(\bx))) - nH(X) &= 
%    \Bigl(n + t\log(|\cX|-1)\Bigr)
%    - \Bigl(n + \frac{n}{2} \log(|\cX|-1)\Bigr)\\
%    &= \bigl(t-\frac{n}{2}\bigr)\log(|\cX|-1).
%  \end{align*}
  Let $P_X$ be the distribution defined by (\ref{eq:3}).  We prove the
  second bound (the proof of the first bound is symmetric).  From
  (\ref{eq:5}) we get
  \begin{align*}
    \Pr_{\bx}\bigl[-\log(P_{X^n}(\bx)) \leq  n \bigl(H(X) - \delta\bigr) \bigr]
    &= 
    \Pr_{\bx}\Bigl[\Bigl(\frac{n}{2}-z(\bx)\Bigr)\log(|\cX|-1)\leq -n\delta\Bigr]\\
    &=
    \Pr_{\bx}\Bigl[\frac{n}{2}-z(\bx) \leq
    -\frac{n\delta}{\log(|\cX|-1)}\Bigr]\\
    &=
    \Pr_{\bx}\Bigl[z(\bx) \geq \frac{n}{2}+
    \frac{n\delta}{\log(|\cX|-1)}\Bigr].
  \end{align*}
  Using Lemma~\ref{lem:bernoullisumestimate} with $s := \lceil
  \frac{n\delta}{\log(|\cX|-1)} \rceil$ (it is easy to check that the
  requirements of Lemma~\ref{lem:bernoullisumestimate} are satisfied)
  we get
  \begin{align}
    \Pr_{\bx}\Bigl[z(\bx) \geq \frac{n}{2}+
    \frac{n\delta}{\log(|\cX|-1)}\Bigr]
    &\geq
    \sum_{k = \lceil\frac{n}{2}\rceil + s}^{\lceil\frac{n}{2}\rceil + 2s-1}
    2^{-n}\binom{n}{k}\nonumber\\
    &>
    \frac{s}{2\sqrt{n}} e^\frac{-8s^2}{n}\nonumber\\
    &\geq
    \frac{\sqrt{n}\delta}{2\log(|\cX|-1)}
    e^{-\frac{8}{n} (\frac{n\delta}{\log(|\cX|-1)}+1)^2}\nonumber\\
    &=
    \frac{\sqrt{n}\delta}{2\log(|\cX|-1)}
    e^{-\frac{8n\delta^2}{\log^2(|\cX|-1)}-\frac{16\delta}{\log(|\cX|-1)} - \frac{8}{n}}\nonumber\\
    &\geq
    \frac{\sqrt{n}\delta}{2\log(|\cX|-1)}
    e^{-\frac{8n\delta^2}{\log^2(|\cX|-1)}-2}\nonumber\\
    &\geq
    \frac{\sqrt{n}\delta}{16\log(|\cX|-1)}
    2^{-\frac{12n\delta^2}{\log^2(|\cX|-1)}}\label{eq:1}\,.
  \end{align}
  Fix now $\delta^* = \frac{\log(|\cX|-1)}{4\sqrt{n}}$.  We consider
  the cases $\delta \geq \delta^*$ and $\delta < \delta^*$ separately.
  First, in case $\delta \geq \delta^*$ equation~(\ref{eq:1}) implies
  \begin{align}\label{eq:2}
    \Pr_{\bx}\Bigl[ z(\bx) \geq \frac{n}{2}+
    \frac{n\delta}{\log(|\cX|-1)}\Bigr]
    > 
    \frac{1}{64}
    2^{-\frac{12n\delta^2}{\log^2(|\cX|-1)}}.
  \end{align}
  On the other hand, if $\delta < \delta^*$ we use (\ref{eq:2}) to get
  \begin{align*}
    \Pr_{\bx}\Bigl[ z(\bx) \geq \frac{n}{2}+\frac{n\delta}{\log(|\cX|-1)}\Bigr]
    \geq
    \Pr_{\bx}\Bigl[ z(\bx) \geq \frac{n}{2}+
    \frac{n\delta^*}{\log(|\cX|-1)}\Bigr]
    \geq
    \frac{1}{64}2^{-\frac{3}{4}} > \frac{1}{110}.
  \end{align*}
  Since $1 \geq 2^{-\frac{12n\delta^2}{\log^2(|\cX|-1)}}$ for all
  $\delta \geq 0$ this finishes the proof.  
\end{proof}

We  now prove that Theorem~\ref{thm:classprodentr} is almost tight.
\begin{theorem} \label{thm:optimalityTwo}
  For any~$\cX$ with~$|\cX|\geq3$ there exists a distribution~$P_X$
  over~$\cX$ such that for~$n \geq 1200$ and~$0\leq \delta \leq
  \frac{\log(|\cX|-1)}{480}$
  \begin{align*}
    \frac{1}{n} \Hmax^\eps(X^n) &\geq H(X)+\delta\ ,\\
    \frac{1}{n} \Hmin^\eps(X^n) &\leq H(X)-\delta\ ,
  \end{align*}
  where~$\eps = \frac{1}{880}2^{-\frac{48n\delta^2}{\log^2(|\cX|-1)}}$.
\end{theorem}
\begin{proof}
  Again, let~$P_X$ be the distribution defined by (\ref{eq:3}).
  First, let~$\cS \subseteq \cX^n$ be the set of values~$\bx$ whose
  probability is at most $P_{X^n}(\bx) \leq
  2^{-n(H(X)+2\delta+10/n)}$, and $\cT \subseteq\cX^n$ be the set of
  values whose probability is at least $P_{X^n}(\bx) \geq
  2^{-n(H(X)-2\delta-10/n)}$.  For both $\cS$ and~$\cT$, according to
  Theorem~\ref{thm:optimalityOne} (one can easily check that
  Theorem~\ref{thm:optimalityOne} can be applied for these
  parameters), the probability that~$\bx$ is in the set is at least
  \begin{align}
    \frac{1}{110} 2^{-\frac{12n(2\delta+10/n)^2}{\log^2(|\cX|-1)}}
    &=
    \frac{1}{110}
    2^{-\frac{12n(4\delta^2+40\delta/n+\frac{100}{n^2})}{\log^2(|\cX|-1)}}\nonumber\\
    &=
    \frac{1}{110}
    2^{-\frac{48n\delta^2}{\log^2(|\cX|-1)}}
    \cdot
    2^{-\frac{480\delta+\frac{1200}n}{\log^2(|\cX|-1)}}\nonumber\\
    &\geq
    \frac{1}{110}
    2^{-\frac{48n\delta^2}{\log^2(|\cX|-1)}}
    \cdot
    2^{-\frac{\log(|\cX|-1)+1}{\log^2(|\cX|-1)}}\nonumber\\
    &\geq
    \frac{1}{440}
    2^{-\frac{48n\delta^2}{\log^2(|\cX|-1)}}=2\eps.\label{eq:9}
  \end{align}
  
  We now prove the bound on~$\Hmax^\eps$.  If $Q\in \cB^\eps(P_{X^n})$
  we have
  \begin{align*}
    \sum_{\bx \in \cS} Q(\bx) 
    \geq  \sum_{\bx \in \cS} P_{X^n}(\bx) - \sum_{\bx \in \cS}|P_{X^n}(\bx) - Q(\bx)|
    \geq \eps,
  \end{align*}
  where the last inequality follows because (\ref{eq:9}) shows
  that~$\sum_{\bx \in \cS} P_{X^n}(\bx)$ is at least~$2\eps$,
  while~$\sum_{\bx \in \cS}|P_{X^n}(\bx) - Q(\bx)|$ is at
  most~$\|P_{X^n}-Q\|_1\leq\eps$.  
  
  We can assume that for all $\bx \in \cS$ we have $Q(\bx) \leq
  2^{-n(H(X)+2\delta+10/n)}$ (otherwise we find a function~$Q(\bx)$
  which is closer to $P_{X^n}$, has the same support, and satisfies
  this) and thus this means that
  \begin{align*}
    |\supp Q|\geq \eps\, 2^{n(H(X)+2\delta+10/n)},
  \end{align*}
  i.e.,
  \begin{align*}
    \log(|\supp Q|)\geq n(H(X)+2\delta+10/n) - \log(1/\eps).
  \end{align*}
  Since 
  \begin{align*}
    \log(1/\eps) 
    &=\frac{48n\delta^2}{\log^2(|\cX|-1)} + \log(880)< n\delta+10
  \end{align*}
  we get that
  \begin{align*}
    \Hmax^\eps &= \min_{Q_{X} \in \cB^\eps(P_{X^n})} 
    \log \bigl|\supp Q_{X} \bigr|\\
    &\geq 
    n(H(X)+2\delta+10/n)  - n\delta-10
    =n(H(X)+\delta)\,.
  \end{align*}
  
  We now come to the bound on the min-entropy~$\Hmin^\eps(X^n)$.  If
  $Q \in \cB^\eps(P_{X^n})$ then $\eps \geq \sum_{\bx \in \cT}
  |P_{X^n}(\bx) - Q(\bx)| \geq \sum_{\bx \in \cT} P_{X^n}(\bx)
  -Q(\bx)$ (where~$\cT$ is as defined above), and together
  with~\eqref{eq:9} this implies
  \begin{align*}
    \sum_{\bx\in\cT} Q(\bx) 
    \geq \sum_{\bx\in\cT} P_{X^n}(\bx)-\eps 
    \geq \sum_{\bx\in\cT} \frac{1}{2}P_{X^n}(\bx),
  \end{align*}
  which implies that for some~$\bx\in\cT$ we have~$Q(\bx) \geq
  2^{-n(H(X)-\delta)}$.
  We therefore get
  \begin{align*}
    \Hmin^\eps(X^n)
    =
    \max_{Q \in \cB^\eps(P_{X^n})}  \min_{\bx \in \cX^n} \log
    \Bigl(\frac{1}{Q(\bx)} \Bigr)
    &\leq
    n(H(X)-\delta)\,.\qedhere
  \end{align*}
\end{proof}
\appendix

\section{Properties of the function $r_t$} \label{sec:rt}

This section lists some properties of the functions $r_t$ defined
by~\eqref{eq:rtdef}, i.e.,
\[
r_t(z) = z^t - t \ln(z) -1 \ .
\]
These properties are used in Section~\ref{sec:smoothprod}.

\begin{lemma} \label{lem:rtincr}
  For any $t \in \bbR$, the function $r_t$ is monotonically increasing
  on the interval $[1, \infty)$.
\end{lemma}

\begin{proof}
  The first derivative of $r_t$ is given by
  \[
    \tfrac{d}{d z} r_t(z) 
  = 
    t z^{t-1} - \tfrac{t}{z} = \tfrac{t}{z} (z^t - 1) \ .
  \]
  The assertion follows because the term on the right hand side is
  nonnegative for any $z \in [1, \infty)$.
\end{proof}

\begin{lemma} \label{lem:rtzz}
  For any $t \in \bbR$ and $z \in (0, \infty)$,
  \[
    r_t(z) \leq r_{|t|}(z+ \tfrac{1}{z}) \ .
  \]
\end{lemma}

\begin{proof}
  Observe first that $r_t(z) = r_{-t}(\frac{1}{z})$.  It thus suffices
  to show that the statement holds for $t \geq 0$. If $z \geq 1$, the
  assertion follows directly from Lemma~\ref{lem:rtincr}. For the case
  where $t \geq 0$ and $z < 1$, let $v := t \ln(1/z)$. Then
  $r_t(\frac{1}{z}) = e^v - v - 1$ and $r_t(z) = e^{-v} + v - 1$.
  Because $v \geq 0$, we have $e^v - e^{-v} \geq 2 v$, which implies
  $r_t(z) \leq r_t(\frac{1}{z})$.  The assertion then follows again
  from Lemma~\ref{lem:rtincr}.
\end{proof}

\begin{lemma} \label{lem:rtconc}
  For any $t \in [-\frac{1}{2}, \frac{1}{2}]$, the function $r_t$ is
  concave on the interval $[4, \infty]$.
\end{lemma}

\begin{proof}
  We show that $\frac{d^2}{dz^2} r_t(z) \leq 0$ for any $z \geq 4$.
  Because $\frac{d^2}{dz^2} r_t(z) = t (t-1) z^{t-2} + \frac{t}{z^2}$,
  this is equivalent to $t(1-t) z^t \geq t$.  It thus suffices to
  verify that
  \[
    z \geq \left(\frac{1}{1-t}\right)^{\frac{1}{t}} \ ,
  \]
  for any $z \geq 4$.  If we substitute~$s := \frac{1}{1-t}$ the right
  hand side of the above expression is $s^{\frac{s}{s-1}}$ (where
  $s\in[\frac23,2]$), whose derivate can be easily seen to be
  non-negative (we use $\ln(s)\leq s-1$), and thus takes its maximum
  at $s=2$, in which case it equals to~$4$.
\end{proof}

\begin{lemma} \label{lem:rtbound}
  For any $z \in [1, \infty)$ and $t \in [-\frac{1}{\log(z)},
  \frac{1}{\log(z)}]$
   \[
     r_t(z) \leq \bigl(1-\ln(2)\bigr) \log^2(z) t^2 \ .
   \]
\end{lemma}

\begin{proof}
  Let $v := t \ln(z)$. Then
  \begin{equation} \label{eq:rtsecond}
    \frac{r_t(z)}{t^2}
  =
    \frac{e^{t \ln(z)} - t \ln(z) -1}{t^2}
  =
    \frac{e^v - v - 1}{v^2} \ln^2(z) \ .
  \end{equation}
  We first show that the term on the right hand side is monotonically
  increasing in $v$, that is,
  \[
    \frac{d}{d v} \frac{e^v - v - 1}{v^2}
  =
    \frac{e^v - 1}{v^2} - 2 \frac{e^v - v -1}{v^3}
  =
    \frac{e^v + 1}{v^2} - \frac{2}{v} \frac{e^v -1}{v^2}
  \geq 
    0 \ .
  \]
  We multiply the last inequality with $v^2e^{-v/2}$ on both sides and
  see that it is equivalent to
  \[
    1 \geq \frac{2}{v}\frac{e^{v/2}-e^{-v/2}}{e^{v/2}+e^{-v/2}} \ ,
  \]
  which holds because, for any $v \in \bbR$, 
  \[
    \Bigl| \frac{e^{v/2}-e^{-v/2}}{e^{v/2}+e^{-v/2}} \Bigr| 
  = 
    | \tanh(\tfrac{v}{2}) | 
  \leq 
    \tfrac{|v|}{2} \ . 
  \]
  
  Hence, in order to find an upper bound on~\eqref{eq:rtsecond}, it is
  sufficient to evaluate the right hand side of~\eqref{eq:rtsecond}
  for the maximum value of $v$. By assumption, we have $v \leq
  \ln(2)$, i.e.,
  \[
    \frac{e^v - v - 1}{v^2} \ln^2(z)
  \leq
    \bigl(1-\ln(2)\bigr) \log^2(z) \ ,
  \]
  which concludes the proof.
\end{proof}

\section{Partial Sums over Binomial Coefficients}\label{sec:binom}
Let
\begin{align}\label{eq:7}
  B_p(k|n) := \binom{n}{k}p^k (1-p)^{n-k} 
\end{align}
be the probability of obtaining~$k$ successes from $n$ independent
Bernoulli trials.  We will also use the binary Kullback-Leibler
distance~$D(q\|p)$, which is defined for arbitrary~$p,q\in[0,1]$ by
\begin{align}\label{eq:8}
  D(q\|p) := q \log\Bigl(\frac{q}{p}\Bigr) + 
  (1-q)\log\Bigl(\frac{1-q}{1-p}\Bigr).
\end{align}

\begin{lemma}\label{lem:kullbackleiblerestimate}
  For $p \geq \frac{1}{2}$, $ \eps \geq 0$, $p + \eps < 1$
  \begin{align*}
    D(p+\eps\| p) \leq \frac{\eps^2}{2\ln(2) p(1-p)}.
  \end{align*}
\end{lemma}
\begin{proof}
  Define the function $f_p(\eps) := D(p+\eps\|p)$.
  Taylor's Theorem states that there exists a $\delta \in [0,\eps]$ such
  that 
  \begin{align}\label{eq:17}
    D(p+\eps\| p) = f_p(\eps) = f_p(0) + f'_p(0)\eps +
    f''_p(0)\frac{\eps^2}{2} + 
    f'''_p(\delta)\frac{\eps^3}{6}.
  \end{align}
  Explicit calculation yields $f_p(0) = f'_p(0) = 0$ and 
  $f''_p(0) = \frac{1}{\ln(2) p (1-p)}$.  Also we get
  \begin{align*}
    f'''_p(\eps) = \frac{2p + 2\eps - 1}{(p+\eps)^2 (p + \eps - 1)^2
      \ln(2)},
  \end{align*}
  which is positive for $\eps > 0$.  Together with~(\ref{eq:17}) this
  gives the lemma.
\end{proof}

\begin{proposition}[Stirling's Approximation]
  For any $n > 0$
  \begin{align}\label{eq:16}
    e^{\frac{1}{12n+1}} 
    <
    \frac{n!\,e^{n}}{\sqrt{2\pi n}\, n ^{n} }
    < 
    e^{\frac{1}{12n}}.
  \end{align}  
% Usual version:
%  \begin{align}\label{eq:15}
%    \sqrt{2\pi}\, n ^{n+\frac12} e^{-n + 1/(12n+1)} 
%    <
%    n!
%    < 
%    \sqrt{2\pi}\, n ^{n+\frac12} e^{-n + 1/(12n)}.
%  \end{align}
\end{proposition}

In the following lemma we are only interested in the lower bound
on~$B_p(k|n)$.  However, the upper bound comes for free.
\begin{lemma}\label{lem:bernoulliestimate}
  For $0 < k < n$ and~$p\in[0,1]$
  \begin{align*}
    e^{-\frac{1}{12k} - \frac{1}{12(n-k)}} <
    B_p(k|n) \sqrt{2\pi \tfrac{k(n-k)}{n}}\, 
    2^{n D(\frac{k}{n}\| p)}  < 1.
  \end{align*}
\end{lemma}
\begin{proof}
  We get
  \begin{multline}
    B_p(k|n) \sqrt{2\pi \tfrac{k(n-k)}{n}} 2^{n
        D(\frac{k}{n}\| p)} \\
    \begin{aligned}
      &= \underbrace{p^k (1-p)^{n-k}}_{2^{(k\log(p) + (n-k)\log(1-p))}}
      \frac{n!}{k!\,(n-k)!}\cdot
      \sqrt{2\pi \tfrac{k(n-k)}{n}}  \cdot
      2^{(k \log(\frac{k}{pn}) + (n-k)\log(\frac{n-k}{(1-p)n}))}\\
      &=
      \sqrt{2\pi\tfrac{k(n-k)}{n}}\cdot 
      \frac{n!}{k!\,(n-k)!}\cdot
      2^{k \log(\frac{k}{n}) + (n-k)\log(\frac{n-k}{n})}\\
      &=
      \sqrt{2\pi\tfrac{k(n-k)}{n}}\cdot
      \frac{n!}{k!\,(n-k)!}\cdot
      \Bigl(\frac{k}{n}\Bigr)^k \cdot\Bigl(\frac{n-k}{n}\Bigr)^{n-k}
  \end{aligned}\\
  =
    \frac{n!\, e^n}{\sqrt{2\pi n}\,n^n}\cdot \frac{\sqrt{2\pi k}\,k^k}{k!\,e^k}
    \cdot\frac{\sqrt{2\pi (n-k)}\,(n-k)^{n-k}}{(n-k)!\, e^{n-k}}.
  \end{multline}
  Using~(\ref{eq:16}) three times we obtain
  \begin{align*}
    B_p(k|n) \sqrt{2\pi \tfrac{k(n-k)}{n}} 2^{n D(\frac{k}{n}\| p)} >
    e^{\frac{1}{12n+1}} e^{-\frac{1}{12k}}e^{-\frac{1}{12(n-k)}} >
    e^{-\frac{1}{12k} - \frac{1}{12(n-k)}}.
  \end{align*}
  Analogously (and since either $\frac{1}{12n} < \frac{1}{12k+1}$ or
  $\frac{1}{12n} < \frac{1}{12(n-k)+1}$)
  \begin{align*}
    B_p(k|n) \sqrt{2\pi \tfrac{k(n-k)}{n}} 2^{n D(\frac{k}{n}\| p)} &< 
      e^{\frac{1}{12n}} e^{-\frac{1}{12k+1}}e^{-\frac{1}{12(n-k)+1}}<1.\qedhere
  \end{align*}
\end{proof}

\begin{corollary}\label{cor:bernoulliestimatesimple}
  For $p\in[\frac12,1]$, and $pn \leq k < n$:
  \begin{align*}
    B_p(k|n) > 
    e^{- \frac{1}{6(n-k)}}
    \sqrt{\tfrac{n}{2\pi k(n-k)}} \, e^{-n\frac{(\frac{k}{n}-p)^2}{2p(1-p)}}.
  \end{align*}
\end{corollary}
\begin{proof}
  From Lemma~\ref{lem:bernoulliestimate} we get
  \begin{align*}
    B_p(k|n) &>
    e^{- \frac{1}{12k} - \frac{1}{12(n-k)}}
    \sqrt{\tfrac{n}{2\pi k(n-k)}} \, 2^{-nD(\frac{k}{n}\|p)}\\
    &\geq
    e^{- \frac{1}{6(n-k)}}
    \sqrt{\frac{n}{2\pi k(n-k)}} \, 2^{-nD(\tfrac{k}{n}\|p)},
  \end{align*}
  where we used $k \geq \frac{n}{2}$.  Using the estimate in
  Lemma~\ref{lem:kullbackleiblerestimate} concludes the proof.
\end{proof}
\begin{lemma}\label{lem:bernoullisumestimate}
  Let $p \in [\frac{1}{2},1]$, $n, s \in \bbN$ such that $pn + 3s \leq
  n$.  Then,
  \begin{align*}
    \sum_{k=\lceil pn\rceil +s}^{\lceil pn\rceil +2s-1} B_{p}(k|n) > 
    \frac{s}{2\sqrt{n}}\, e^{-\frac{2s^2}{n p(1-p)}}.
  \end{align*}
\end{lemma}
\begin{proof}
  Clearly, $\frac{n}{k (n-k)} \geq\frac{4}{n}$, for all values of~$n$
  and~$k$ in the sum.  Since $k < pn + 2s$ for all values in the above
  sum we get $\frac{k}{n}-p < \frac{2s}{n}$, and also we see that $n-k
  \geq s$.  Using this together with
  Corollary~\ref{cor:bernoulliestimatesimple} thus implies for all~$k$
  of interest
  \begin{align*}
    B_p(k|n) &> 
    e^{- \frac{1}{6s}}
    \sqrt{\frac{2}{\pi n}} \, e^{-n\frac{(\frac{2s}{n})^2}{2p(1-p)}} \\
    & >
    \frac{1}{2 \sqrt{n}} \, e^{-\frac{2s^2}{np(1-p)}}\,.
  \end{align*}
  Since there are $s$ summands we get the lemma.
\end{proof}

%% \bibliography{qkd}

%% \bibliographystyle{alpha}

\end{document}